\documentclass[12pt]{article}

\usepackage[super,comma]{natbib}
\usepackage{graphicx}

\usepackage{times}

\title{Cosmic dawn as a dark matter detector}

\author
{Rennan Barkana,$^{1}$\\
\\
\normalsize{$^{1}$Raymond and Beverly Sackler School
of Physics and Astronomy,}\\
\normalsize{Tel Aviv University, Tel Aviv 69978, Israel}\\
}

\begin{document} 


\baselineskip24pt


\maketitle 

{\bf 

Models and simulations predict that the cosmic radio spectrum should
show a strong absorption signal corresponding to the 21-cm transition
of atomic hydrogen around redshift 20, due to Lyman-$\alpha$ radiation
from some of the earliest stars\cite{Madau,21cmRev,21cmRev2}. An
international effort is underway to detect cosmic dawn by observing
this 21-cm signal, either its sky-averaged spectrum\cite{Judd} or maps
of 21-cm fluctuations using radio interferometers\cite{HERA,SKA}. Here
we show that a detection of this signal can potentially yield not only
a discovery of the first stars, but also a direct detection of dark
matter if it scatters with baryons. A stronger absorption signal than
expected, as strongly indicated by the first observational detection
of the global 21-cm spectrum\cite{Judd}, implies excess cooling of the
cosmic gas, which can only be plausibly induced by an interaction with
dark matter. The observed signal implies that the mass of the dark
matter particle is below a few GeV (there is no lower limit), and that
it scatters off baryons with a cross-section $\sigma$ of at least
$10^{-21}$~cm$^2$ at a relative velocity $v$ of 1~km/s (corresponding
to $10^{-43}$~cm$^2$ at the speed of light, for a $\sigma(v) \propto
v^{-4}$ model). The measured signal also directly confirms that the
dark matter is highly non-relativistic and rather cold; primordial
velocities in the range corresponding to warm dark matter models are
potentially detectable. These results transform 21-cm cosmology into a
unique dark matter probe, strengthened by the prediction that the
root-mean-square 21-cm fluctuation at cosmic dawn is likely to be an
order of magnitude larger than previously expected.}

An excess 21-cm absorption signal is a clear sign of baryon - dark
matter (b-DM) scattering, since alternative explanations are
untenable. In general, the intensity of the 21-cm signal is expressed
as the observed mean brightness temperature relative to the cosmic
microwave background (CMB), and is given
by\cite{Madau} \begin{equation} T_{21} = 26.8\, x_{\rm H\, I}\,
  \frac{\rho_{\rm g}}{\bar{\rho}_{\rm g}} \left( \frac{\Omega_{\rm b}
    h} {0.0327} \right) \left(\frac{\Omega_{\rm m}}{0.307}\right)^
       {-1/2} \left( \frac{1+z} {10} \right)^{1/2} \left(
       \frac{T_\mathrm{S}-T_\mathrm{CMB}} {T_\mathrm{S}} \right)\,
       \mathrm{mK}\ ,
             \label{e:Tb} \end{equation}
where $x_{\rm H\, I}$ is the mean mass fraction of hydrogen that is
neutral (i.e., not ionized), $\rho_{\rm g}$ is the gas density and
$\bar{\rho}_{\rm g}$ its cosmic mean value, $\Omega_{\rm m}$ and
$\Omega_{\rm b}$ are the cosmic mean densities of matter and of
baryons, respectively, in units of the critical density, $h$ is the
Hubble parameter in units of $100\, \mbox{ km s}^{-1}\,
\mbox{Mpc}^{-1}$, $z$ is the redshift [corresponding to an observed
wavelength of 21$\times (1+z)$~cm and an observed frequency of
$1420/(1+z)$~MHz], $T_\mathrm{CMB}=2.725\times(1+z)$ is the CMB
temperature at $z$, and $T_\mathrm{S}$ is the spin temperature of
hydrogen at $z$. The latter quantity is an effective temperature that
describes the relative abundances of the ground and excited states of
the hyperfine splitting (spin-flip transition) of the hydrogen atom;
in the absence of astrophysical radiation, this temperature is set by
collisions of the hydrogen atoms with other atoms and scattering of
CMB photons\cite{Purcell}, and therefore $T_\mathrm{gas} \le
T_\mathrm{S} \le T_\mathrm{CMB}$, where $T_\mathrm{gas}$ is the
(kinetic) gas temperature.

Observations of the 21-cm line can be used to probe density
fluctuations\cite{Hogan}, cosmic reionization\cite{Scott}, and X-ray
heating\cite{Madau,Fur06,Cold}, but the earliest observable milestone
during cosmic dawn is an absorption
signal\cite{Madau,21cmRev,21cmRev2}, expected once stellar
Lyman-$\alpha$ photons indirectly couple $T_\mathrm{S}$ to
$T_\mathrm{gas}$ via the Wouthuysen-Field effect\cite{Wout,Field}. The
first detection of a cosmic 21-cm signal is the EDGES detection of the
global spectrum from cosmic dawn\cite{Judd}, which found a brightness
temperature $T_{21} = -500 \pm 200$~mK corresponding to peak
absorption at frequency $\nu \sim 78$~MHz ($z = 17.2$). If confirmed,
this signal (which is well below $-209$~mK, the strongest possible
absorption at this frequency under standard expectations) cannot be
explained within the standard paradigm, even if one allows for exotic
astrophysics (see Supplementary Information section S1). Basic
thermodynamics suggests that it is easy to heat the cosmic gas but
difficult to cool it. The extra cooling indicated by the data is only
possible through the interaction of the baryons with something that is
even colder.

The only known cosmic constituent that can be colder than the early
cosmic gas is the dark matter. The reason for this is that DM is
assumed to interact with itself and with baryons only gravitationally,
and thus it is expected to thermally decouple in the very early
Universe and cool down thereafter (particularly fast if it is
non-relativistic early on, as in the case of cold dark matter). Any
significant electrodynamic or nuclear interactions of the DM would be
inconsistent with the great observational successes of standard
cosmology, including Big-Bang nucleosynthesis, CMB observations, and
the formation and distribution of galaxies. However, weak
non-gravitational interactions are possible. There is a wide array of
possibilities for how the strength of such an interaction might vary
with temperature, or more specifically with the relative velocity
between the baryon and the DM particle that it scatters with.  A
crucial point is that cosmic dawn presents unique physical conditions
that can probe a range of parameters that are encountered no-where
else. In particular, at this time the cosmic gas is at its coldest, as
it is hotter before (due to its remnant thermal energy from the Big
Bang), and hotter afterwards (due to X-rays and other heating
radiation from astrophysical objects). Thus, if b-DM scattering
happens to be strongest at low relative velocities, its effect might
show up only at cosmic dawn.

The cross-section for b-DM collisions is normally expressed at a
relative velocity equal to the speed of light (we denote this
$\sigma_c$), but we express it at 1~km/s ($\sigma_1$), which is close
to the typical velocities during cosmic dawn (though in some models
they reach below 0.1~km/s). We adopt a $v^{-4}$ dependence that has
often been used to illustrate the case of a strongly increasing
cross-section at low velocities, i.e.,
\begin{equation}
  \sigma(v) = \sigma_c \left( \frac{v}{c} \right)^{-4} = \sigma_1
  \left( \frac{v}{1~{\rm km/s }} \right)^{-4}\ .
  \label{e:sigv}
\end{equation}
Such a velocity dependence would arise naturally in the case of DM
millicharge, i.e., if the DM has a small electric charge and the
interaction is via Rutherford/Coulomb scattering\cite{Davidson,Dubovsky,McDermott}.

We calculate the thermal evolution by following the exchange of energy
and momentum between the baryons and the DM\cite{Marc,Joe,Kovetz}. A
major role is played by the relative velocity of baryons and DM left
over after cosmic recombination\cite{Kovetz}. This long-recognized
velocity remnant\cite{SZ} arises from the fact that the DM motion is
determined by gravity while the baryons had been scattering rapidly
off the CMB photons and moving along with them in their acoustic
oscillations. This relative velocity (also termed the streaming
velocity) has received attention recently due to its effect on early
galaxy formation\cite{TH10}, which may produce an observable 21-cm
signature\cite{Dalal,Eli}; however, the b-DM scattering that we
consider here depends directly on the velocities, and their effect on
galaxies does not play a role.

The b-DM relative velocity varies spatially (Fig.~1), with a
large-scale pattern of coherent regions\cite{TH10} of size $\sim
100$~Mpc. Since the root-mean-square (r.m.s.) velocity is supersonic
(going from a Mach number of $\sim 5$ right after recombination to
$\sim 2$ when the gas thermally decouples from the CMB), and the
scattering cross-section (eq.~\ref{e:sigv}) varies strongly with the
relative velocity, the evolution in each region depends on the local
value of the initial velocity\cite{Kovetz}. A higher relative velocity
usually implies less cooling, as the result of the interplay of two
factors: the scattering is weaker (at least until the relative
velocity is dissipated away by the b-DM scattering), and the kinetic
energy of the relative velocity is partially transferred into heating
of the baryons. The dependence on velocity yields order unity 21-cm
fluctuations (Fig.~1), which we average over (using the Maxwellian
distribution of the magnitude of the relative velocity\cite{TH10}) in
order to find the global 21-cm signal.

\begin{figure*}[]
\centering
\includegraphics[width=3.0in]{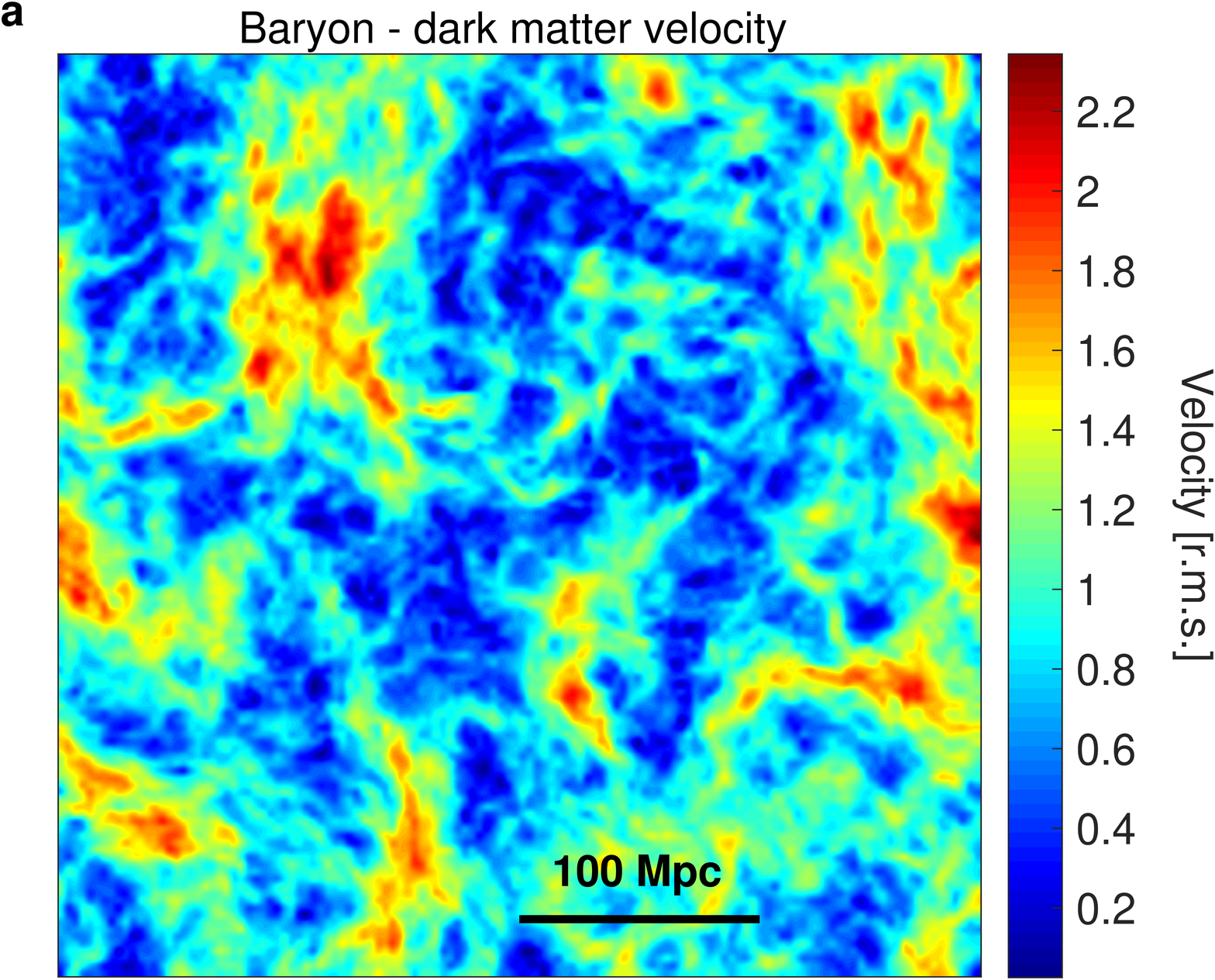}
~ ~ ~ ~
\includegraphics[width=2.9in]{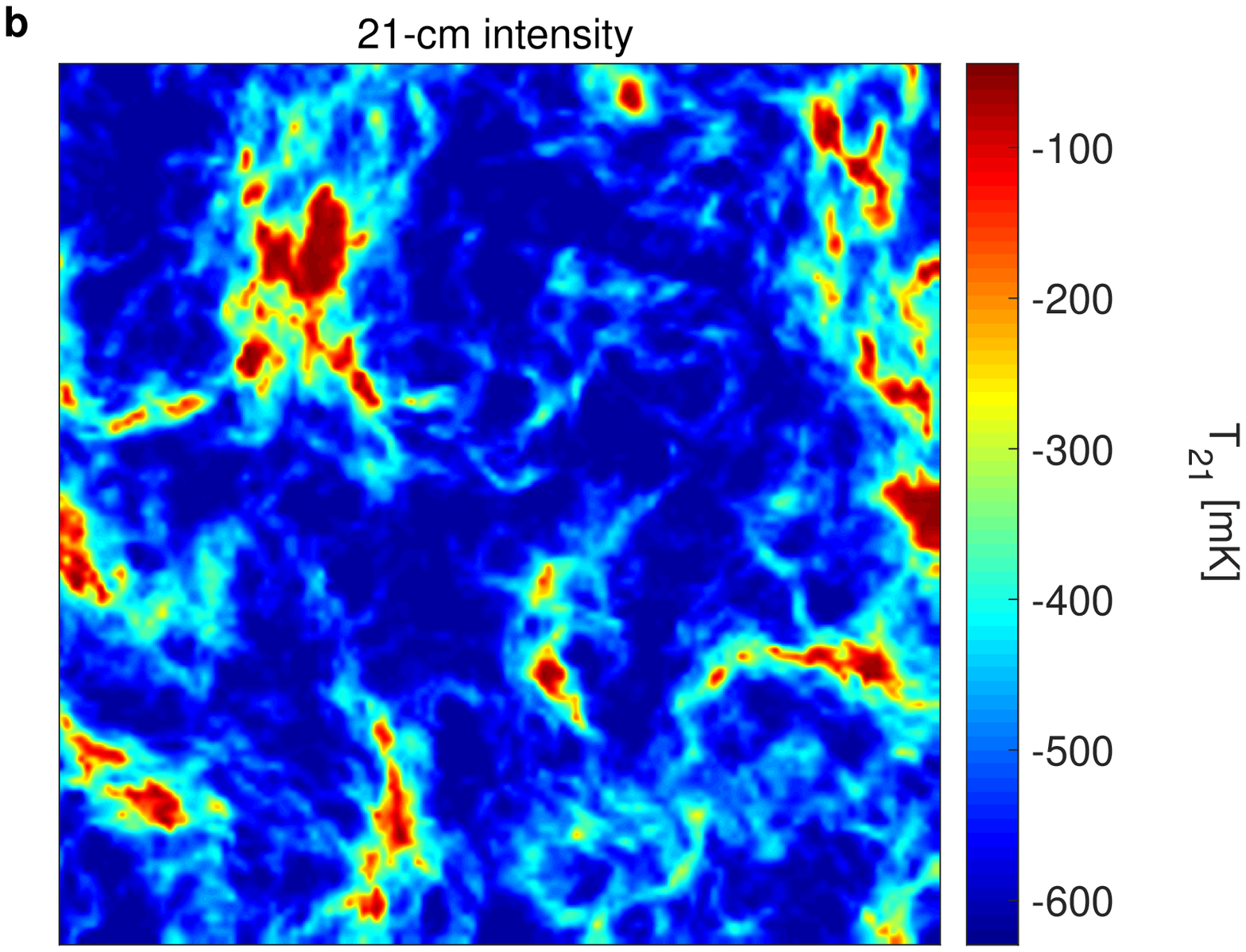}
\caption{{\bf Image of the 21-cm brightness temperature in a model
    with baryon - dark matter scattering.} The 21-cm brightness
  temperature (in units of mK) is shown (Right panel) in a
  two-dimensional slice (thickness = 3~Mpc) of a simulated volume of
  the universe 384~Mpc (all lengths comoving) on a side. We consider
  $z=17$ ($\nu=78.9$~MHz) at which this model (with $\sigma_1 =
  8\times10^{-20}$~cm$^{2}$ and DM particle mass $m_{\chi}=0.3$~GeV)
  reaches its maximum global 21-cm absorption depth of -504~mK
  (roughly matching the most likely observed value\cite{Judd}). The
  spatial 21-cm pattern is determined by the relative b-DM velocity
  left over from early cosmic evolution prior to recombination; its
  distribution (in a randomly-generated example, assuming adiabatic
  initial density fluctuations) is also shown (Left panel), in units
  of its r.m.s.\ value of 29~km/s at $1+z=1010$. We note that we
  simply treat the baryons as equal-mass particles of mass 1.22 times
  the proton mass (which is the mean molecular mass of neutral
  primordial gas). Thus, eq.~\ref{e:sigv} represents the scattering
  cross-section of DM with an average baryon. In reality, the
  treatment of helium is likely to be complicated and highly
  model-dependent\cite{Marc}. We start our calculation at kinematic
  decoupling ($1+z = 1010$) as in previous calculations\cite{Kovetz};
  we have checked that starting earlier would not significantly impact
  our results at lower redshifts. In addition to b-DM scattering, we
  include at each redshift spatially-uniform backgrounds of
  astrophysical radiation of the three types important in 21-cm
  cosmology (Lyman-$\alpha$ photons, X-rays, and ionizing photons). We
  use the volume-averaged values from a semi-numerical
  simulation\cite{Cold,GlobalStudy} with astrophysical parameters
  chosen to illustrate an absorption dip at a redshift consistent with
  the observed signal; the relevant parameters are that star-formation
  occurs only in halos that allow atomic cooling and with an
  efficiency $f_* = 1.58\%$, and X-rays normalized based on
  low-redshift observations are emitted with a soft power-law
  spectrum. The astrophysical radiation fields are actually expected
  to vary spatially, leading to 21-cm fluctuations during cosmic dawn
  due to Lyman-$\alpha$ fluctuations\cite{zCut} and X-ray heating
  fluctuations\cite{Xrays}; these fluctuations are significant and
  potentially observable, but we neglect them here due to the much
  larger fluctuations resulting from b-DM scattering. We assume the
  known values of the cosmological parameters\cite{Planck}.}
\end{figure*}

Such a calculation has been done previously\cite{Kovetz} only during
the dark ages, before the formation of any astrophysical sources. In
that regime, b-DM scattering can yield significant absorption (see the
$\nu < 33$~MHz portion of Fig.~2; only absorption as low as -70~mK has
been previously considered), but this is unlikely to be observed in
the near future since such low-frequency observations would be very
difficult both due to ionospheric distortions and since the galactic
synchrotron foreground\cite{21cmRev} at $\nu = 20$~MHz is $\sim 40$
times stronger than at $\nu = 80$~MHz. A purely cosmological signal
would disappear after the dark ages (at $\nu \sim 50$~MHz), since the
expansion of the universe and the cooling of the gas make coupling of
the 21-cm line to $T_\mathrm{gas}$ (through atomic collisions) less
effective than the coupling to the CMB. This drives $T_\mathrm{S}
\rightarrow T_\mathrm{CMB}$ and eliminates the 21-cm
signal.

Ly$\alpha$ coupling during cosmic dawn\cite{Madau} should reveal the
presence of the cold gas via a strong 21-cm absorption dip. Only a
combination of both b-DM scattering and the formation of the first
stars during cosmic dawn can explain the strong absorption feature
measured by the EDGES experiment (Fig.~2). The existence and shape of
the absorption dip implies early astrophysically-generated
Lyman-$\alpha$ and X-ray radiation backgrounds, consistent with the
absence of a strong absorption signal at higher frequencies (see
Supplementary Information section S2), while the unexpectedly large
depth of the absorption indicates cosmic gas that had been
substantially cooled by b-DM scattering.

\begin{figure*}[]
\centering
\includegraphics[width=4.8in]{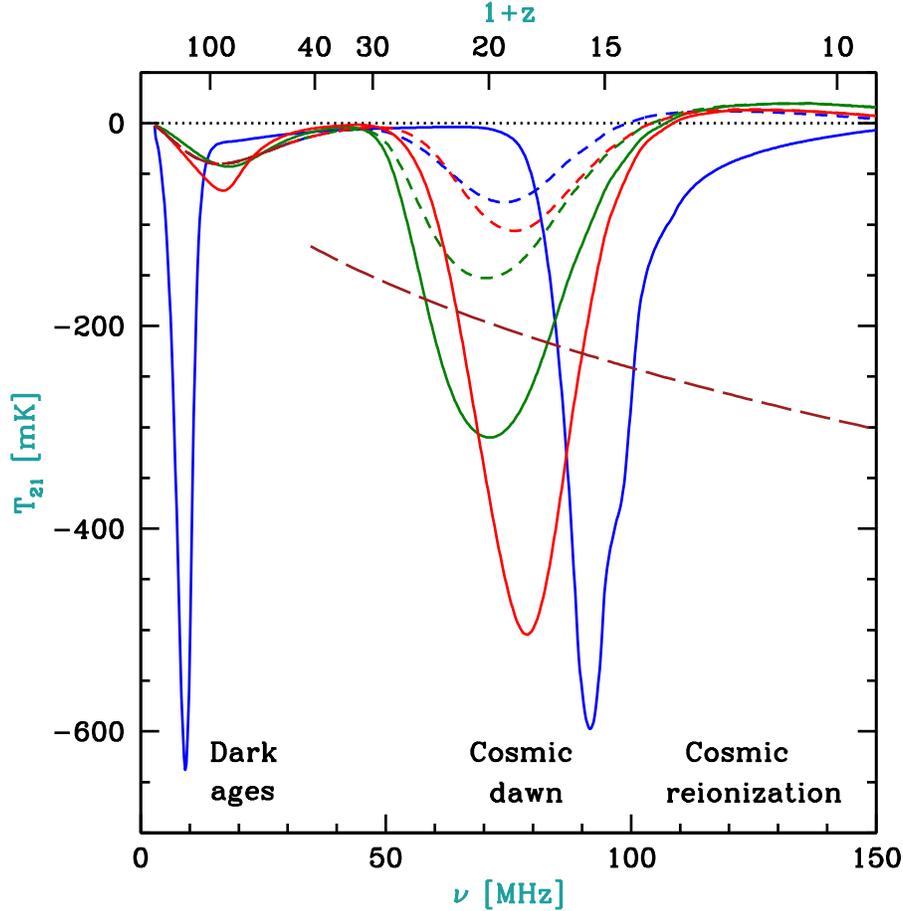}
\caption{{\bf Global 21-cm signal in models with b-DM scattering.}
  The globally-averaged 21-cm brightness temperature (in units of mK)
  is shown at an observed frequency $\nu$ (in MHz), with the
  corresponding value of one plus the redshift shown at the top. We
  chart some of the space of possible 21-cm signals with three
  illustrative models (solid curves): Cross-section $\sigma_1 =
  8\times10^{-20}$~cm$^{2}$ and DM particle mass $m_{\chi}=0.3$~GeV
  (red; roughly matching the most likely observed value\cite{Judd} of
  the peak absorption), $\sigma_1 = 3\times10^{-19}$~cm$^{2}$ and
  $m_{\chi}=2$~GeV (green), and $\sigma_1 = 1\times10^{-18}$~cm$^{2}$
  and $m_{\chi}=0.01$~GeV (blue). These models assume various
  astrophysical parameters; the corresponding 21-cm signals in the
  absence of b-DM scattering are shown as short-dashed curves, red
  (same model as assumed in Fig.~1), green (same model except that the
  efficiency of production of Ly$\alpha$ photons is 10 times higher),
  and blue (same except that $f_* = 0.5\%$, star formation occurs in
  halos that allow molecular cooling, the X-ray efficiency is 4 times
  higher, and the X-ray spectrum extends down to 0.1~keV instead of
  our standard 0.2~keV). Also shown for comparison (long-dashed brown)
  is the standard dark ages prediction with no b-DM scattering (at
  $\nu < 33$~MHz; matches all the short-dashed curves in this range)
  and (at $\nu > 33$~MHz) the lowest global 21-cm signal at each
  redshift that is possible with no b-DM scattering (regardless of the
  astrophysical parameters). We note that in most of the model space,
  the cosmic dawn absorption dip is well fitted by a simple Gaussian
  (though not for the fairly exotic blue solid curve). The current
  measurement\cite{Judd} suggests a somewhat different
  flattened-Gaussian shape, although the significance of this
  difference is unclear given the systematic noise.}
\end{figure*}

The observed 21-cm signal can be explained with a wide range of DM
properties, in terms of the particle masse and b-DM scattering
cross-section (Fig.~3; also see Supplementary Information section
S3). The DM particle mass must be lighter than a few GeV, which 
is much lighter than expected for a weakly-interacting
massive particle (WIMP), but there is no lower limit on the mass, even
down to $m_{\chi} \sim 10^{-31}$~GeV as in ultra-light fuzzy
DM\cite{fuzzy}. A minimum scattering cross-section is required (nearly
independent of the particle mass) of $\sigma_1 > 10^{-21}$~cm$^2$,
which for the $v^{-4}$ model (eq.~\ref{e:sigv}) corresponds to
$\sigma_c > 10^{-43}$~cm$^2$. There is no maximum cross-section, so
that in terms of particle mass and cross-section, cosmic dawn is
sensitive to an enormous range of DM parameter space, most of which is
unavailable to other current or future probes (see Supplementary
Information section S4).

\begin{figure*}[]
\centering
\includegraphics[width=6.3in]{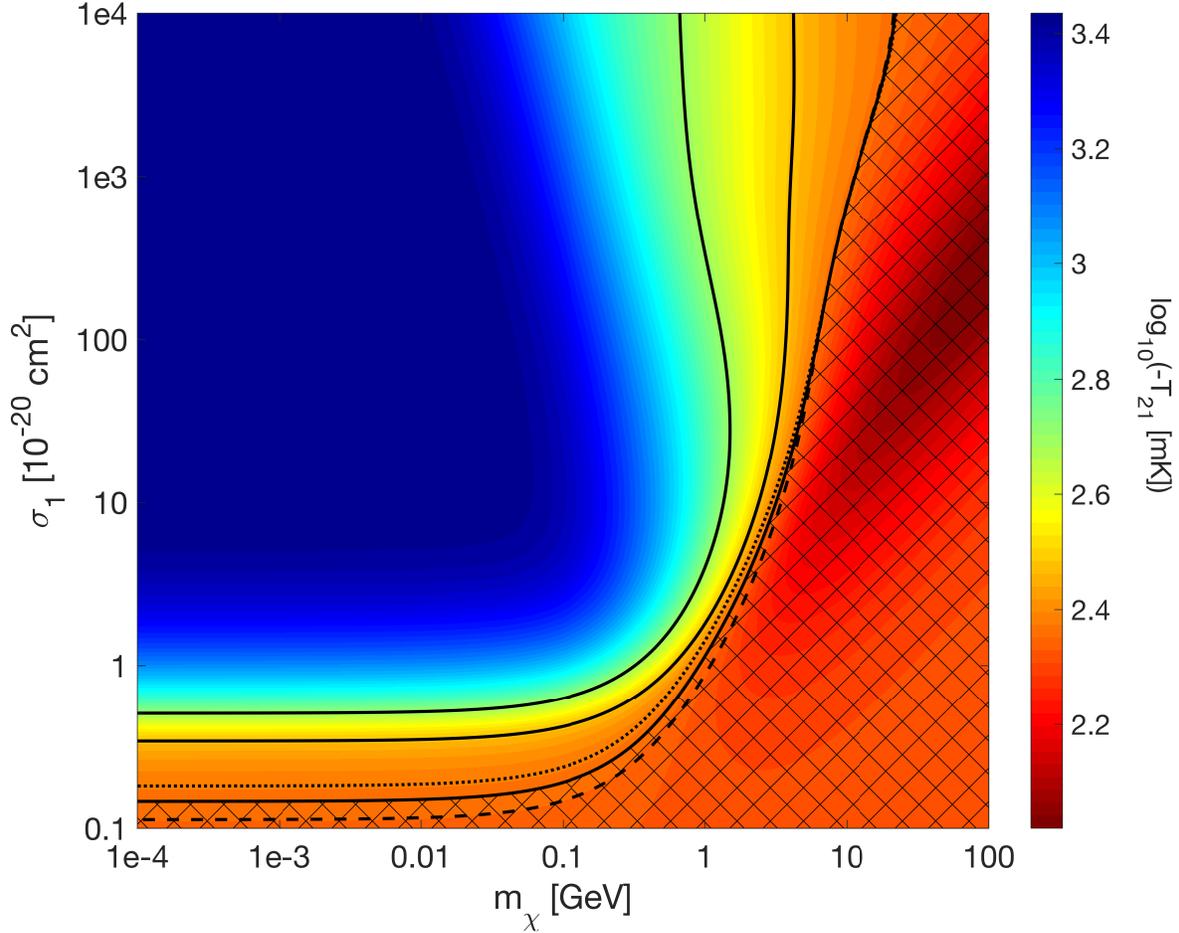}
\caption{{\bf Constraints on dark matter properties using cosmic
    dawn.}  We show the minimum possible 21-cm brightness temperature
  $T_{21}$ (shown as $\log_{10}$ of its absolute value) at $z=17$
  ($\nu = 78.9$~MHz), regardless of astrophysics (i.e., assuming
  saturated Lyman-$\alpha$ coupling and no X-ray heating), as a
  function of the DM particle mass $m_{\chi}$ and the b-DM
  cross-section $\sigma_1$ (eq.~\ref{e:sigv}). Also shown (solid black
  curves) are contours corresponding to $T_{21}$ being more negative
  than the strongest possible absorption depth without b-DM
  scattering, by $10\%$, $50\%$, or $100\%$ (from right to left). The
  hatched region is excluded assuming only that the data\cite{Judd}
  has established absorption by at least $10\%$ more than the minimum
  no-interaction value (at $z=17$ the minimum no-interaction value
  equals -210~mK, or 2.32 on the plotted scale); this implies
  $\sigma_1 > 1.5 \times 10^{-21}$~cm$^2$ and $m_{\chi} < 23$~GeV
  (though any $m_{\chi}$ above a few GeV requires high $\sigma_1$, a
  parameter combination that is likely in conflict with other
  constraints; see Supplementary Information section S4). If the data
  firmly establishes absorption by at least $50\%$ more than the
  no-interaction value, this implies $\sigma_1 > 3.6 \times
  10^{-21}$~cm$^2$ and $m_{\chi} < 3.5$~GeV. We also illustrate the
  redshift dependence of these limits via the corresponding $10\%$
  contours at $z=14$ (dashed) and $z=20$ (dotted). This plot focuses
  on the global 21-cm signal, but note that the corresponding 21-cm
  fluctuations are also in most cases much larger than expected in the
  absence of b-DM scattering.}
\end{figure*}


While we have assumed that $\sigma(v) \propto v^{-4}$, the velocity
dependence of the cross-section can be explored with future 21-cm
data. Further global 21-cm measurements may help, but are unlikely to
resolve the degeneracies among the DM parameters (cross-section
amplitude, velocity dependence, particle mass) and the astrophysical
parameters (the X-ray spectrum, and the normalization and redshift
evolution of the various radiation backgrounds as determined by
parameters related to galaxy formation). Detailed measurements of
21-cm fluctuations, including the 21-cm power spectrum, would provide
far more information. For example, the level of the fluctuations
caused by the spatially varying relative b-DM velocity (Fig.~1)
directly depends on how rapidly the scattering cross-section varies
with velocity. More futuristic would be a comparison with a
measurement of the signal during the dark ages (Fig~2), which would
involve somewhat different velocities and would provide constraints
independent of astrophysics; however, even the proposed DARE
satellite\cite{DARE} only goes down to $\nu=40$~MHz, while the
predicted signal from the dark ages is at $10-30$~MHz.

The observed signal also places the first direct limit on
early-universe scenarios in which the dark matter is not completely
cold, i.e., has a relic thermal velocity. The dark matter must be
colder than the baryons in order to cool them, so if we demand that it
be colder than $T_{\chi}^{20}$ at $z=20$, its r.m.s.\ velocity at
$1+z=1010$ must be $v_{\rm rms}^{1010} <
20$~km/s~$\sqrt{T_{\chi}^{20}/[10~{\rm K}]}\sqrt{1~{\rm
    GeV}/m_\chi}$. In addition, in order for the DM to significantly
cool the gas at cosmic dawn without heavily disrupting the CMB, it
must reach a thermal velocity as low as a few km/s at $z=20$, which
puts an upper limit (independent of $m_\chi$) of $\sim 150$~km/s on
$v_{\rm rms}^{1010}$. Current limits on warm dark matter\cite{WDM}
allow models with $m_\chi \sim 3$~keV which have a corresponding
$v_{\rm rms}^{1010} \sim 10$~km/s. Such thermal motion evades these
upper limits but is comparable to the b-DM relative velocity which
dominates the 21-cm pattern (Fig.~1), and should thus produce
significant observable effects.

Astronomical testing and possible confirmation of the observed
signal\cite{Judd} and of its interpretation in terms of b-DM
scattering is likely to proceed in two tracks. Other global 21-cm
experiments such as SARAS\cite{SARAS} and LEDA\cite{LEDA} should try
to confirm the measured global signal. Meanwhile, upcoming 21-cm
fluctuation experiments aimed at cosmic dawn will provide a definitive
test, since the expected 21-cm intensity pattern should clearly
display a transformed version of the spatial pattern expected for the
relative b-DM velocity (Fig.~1). Experiments such as HERA\cite{HERA}
and the Square Kilometer Array (SKA)\cite{SKA} should easily be able
to measure the corresponding 21-cm power spectrum, since the
r.m.s.\ fluctuation in an illustrative model with b-DM scattering
(Fig.~1) is 140~mK, compared to a previously expected value of at most
$\sim 20$~mK. The large spatial scale of the fluctuation pattern (of
order 100 comoving Mpc, which corresponds to half a degree) will also
make it easier to observe, as it means that high angular resolution is
not needed. As in the case of the galaxy-driven effect of the b-DM
relative velocity\cite{TH10,Dalal,Eli}, the power spectrum should show
a strong baryon acoustic oscillation (BAO) signature, as the velocity
arises in part from the participation of the baryons in the sound
waves of the primordial baryon-photon fluid; indeed, the BAO
oscillations versus wavenumber should be of order unity and have high
peaks at $k \sim 0.03$/Mpc and 0.07/Mpc, as in the power spectrum of
the relative velocity itself\cite{TH10}. A precision measurement at
cosmic dawn of the BAO scale (and thus the angular diameter distance)
would be a useful cosmological tool to add to current constraints
based on similar measurements from low-redshift galaxy
clustering\cite{BAO}. If most stars form in galactic halos below $\sim
10^7 M_\odot$, then star formation should follow the same BAO
pattern\cite{TH10,Dalal,Eli} and be strongly anticorrelated with the
baryon temperature.

The striking predicted spatial pattern (Fig.~1) should make 21-cm
imaging during cosmic dawn possible for the SKA given its expected
sensitivity\cite{SKA}. The expected probability distribution function
(PDF) of the 21-cm intensity is a transformed Maxwellian, which is
highly asymmetric; imaging could directly verify this striking and
previously-unexpected non-Gaussianity. The presence of DM has
historically been inferred by assuming the validity of General
Relativity on galactic and cosmological scales; confirmation of the
existence of DM would thus also constitute another triumph for
Einstein.

\noindent {\bf Supplementary Information} is linked to the online
version of the paper at www.nature.com/nature.

\noindent {\bf Acknowledgments} I am grateful to Judd Bowman for
alerting me to possible indications of very deep absorption in the
EDGES results, which inspired this work. This project/publication was
made possible through the support of a grant from the John Templeton
Foundation. The opinions expressed in this publication are those of
the author and do not necessarily reflect the views of the John
Templeton Foundation.

\noindent {\bf Author Information} Reprints and permissions
information is available at www.nature.com/reprints. Correspondence
and requests for materials should be addressed to
R.B.\ (barkana@tau.ac.il).

\begin{center}
 {\bf \Large Supplementary Information}
\end{center}

\noindent {\bf S1. Strongest possible absorption without baryon - dark
  matter scattering}

A measurement of stronger than expected absorption is a smoking gun
for dark matter, since such absorption cannot be produced without b-DM
scattering. In the standard picture, the best-case scenario for
producing strong 21-cm absorption is to assume no reionization (i.e.,
$x_{\rm H\, I} = 1$ in eq.~\ref{e:Tb}), saturated coupling (i.e.,
$T_\mathrm{S} = T_\mathrm{gas}$), and no astrophysical heating. In
this case, the gas at high redshifts is colder than the CMB since its
adiabatic cooling is faster. However, the baryons are thermally
coupled to the CMB via Compton heating until $z \sim 150$. This
well-understood physics yields\cite{hyrec} a strongest possible
absorption signal (regardless of the uncertain astrophysics at high
redshift) of $T_{21} = -209$~mK at $\nu = 78$~MHz. We note that
$-209$~mK is the maximum possible absorption, i.e., it is an extreme
value (in the standard case without b-DM scattering) that would not be
considered particularly likely; models with various astrophysical
parameter values\cite{GlobalStudy} predict $T_{21}$ values at $\nu =
78$~MHz that range from $-209$~mK up to positive values, with most
falling between -150~mK and -50~mK. More generally, the lowest global
21-cm signal at each frequency that would be possible with no b-DM
scattering (regardless of the details of high-redshift astrophysics)
is shown (at $\nu > 33$~MHz) by the brown dashed curve in Fig.~2.

We can consider various ideas for increasing the absorption without
b-DM interactions. Fluctuations in the gas density $\rho_{\rm g}$
affect the 21-cm signal, as the absorption strength $\propto \rho_{\rm
  g}$. However, adiabatic heating with $T_\mathrm{gas} \propto
\rho_{\rm g}^{2/3}$ counteracts this and leads to only a small
increase in the absorption in overdense regions, while in the voids
these factors combine to weaken the overall absorption. Actually,
linear fluctuations are symmetrical and cancel out when averaged
globally over the overdense and underdense regions. To change the
observed signal, significantly non-linear fluctuations are needed. The
regime of mildly non-linear density fluctuations is well understood,
as it corresponds to the sheets and filaments of the cosmic web that
successfully explain\cite{McQuinn} the observed properties of the
Lyman-$\alpha$ forest at $z=2-5$. The Universe at cosmic dawn is
expected to be much more homogeneous, with the density fluctuations
less non-linear, since gravity had not had as much time to drive the
growth of fluctuations. Nevertheless, even if we were to assume that
somehow the density fluctuations corresponding to the Lyman-$\alpha$
forest were already in place at $z \sim 20$, this still would not
produce a deeper absorption signal. To check this quantitatively, we
assume the best case of $T_\mathrm{S} = T_\mathrm{gas}$ and adiabatic
heating/cooling, and average the 21-cm brightness temperature over the
density distribution at $z=2-6$ in simulations that match
Lyman-$\alpha$ observations\cite{MHR00}. The result we find is a
weaker average absorption than would occur in the absence of any
density fluctuations. More evidence that density fluctuations do not
produce unusual absorption comes from numerical simulations of the
Universe during cosmic dawn; these have been run on a variety of
volumes and resolutions\cite{Ross17,McQuinn2,Benoit}, and none have
predicted a stronger globally-averaged absorption signal than the
simple limit shown in Fig.~2.

While under standard cosmology the total gas fraction within
virialized halos at $z=20$ is expected to be below $1\%$, we can
consider an exotic scenario where unexpectedly large density
fluctuations on small scales would produce a much larger abundance of
minihalos. This also would not produce more absorption. The lowest
$T_\mathrm{gas}$ at $z=20$ in the standard scenario is $\sim 9$~K at
the cosmic mean density. As the gas adiabatically heats, it reaches
the CMB temperature (57~K at $z=20$) at a modest overdensity of
16. After that point it contributes extra emission, not
absorption. When the gas enters a virialized halo, it likely
shock-heats. If it cools efficiently, primordial cooling via molecular
hydrogen is only effective down to temperatures of a few hundred K,
and in any case, efficient cooling likely leads to star formation and
even more heating. We also note that the 21-cm optical depth of the
coldest-possible gas (without b-DM scattering) is $\tau_{21} \sim
10\%$ at the mean density at $z=20$; this varies as $\tau_{21} \propto
\rho_{\rm g}/T_\mathrm{gas} \propto \rho_{\rm g}^{1/3}$ assuming
adiabatic evolution. This means that only very dense gas (inside
virialized halos) can be optically thick.

Another possibility is to change the residual electron fraction after
recombination, which determines the rate of Compton heating that keeps
the gas close to the CMB temperature until $z \sim 150$. To produce
unusually strong absorption, for instance $T_{21} = -300$~mK at $\nu =
77$~MHz, the gas would need to thermally decouple at a $1+z$ that is
higher by a factor of 1.4, which would happen if the residual ionized
fraction were lower than expected by about a factor of 4. Before
cosmic recombination, the gas is strongly coupled to the CMB and
cannot cluster, so it would likely be unaffected even by exotic
physics such as unusually strong DM clumping. After the freeze-out at
the end of cosmic recombination, the recombination time continues to
go up as $1/\rho_{\rm g}$ so that the residual electron fraction only
changes slowly with time, and is only weakly dependent on density (in
part because the recombination coefficient declines with temperature,
and the latter rises with density). It is difficult to imagine
something that could lower the mean residual electron fraction by a
large factor.

More generally, it would be difficult to substantially change the
physics involved in cosmic recombination, the basic cosmological
parameters, or the cosmic expansion history. These inputs are strongly
constrained by the success of standard cosmology in fitting
observations of the CMB plus low-redshift observations. Possible ideas
for exotic astrophysics or physics such as unexpected populations of
stars or black holes, or dark matter annihilation or decay, also fail
to strengthen the absorption. Such scenarios would generate extra UV,
X-ray or gamma-ray radiation, which would generate more heating as
well as more ionization (which would boost the Compton heating of the
gas and lower $x_{\rm H\, I}$); also, Lyman-$\alpha$ coupling cannot
get any stronger than the saturated coupling limit ($T_\mathrm{S} =
T_\mathrm{gas}$) that we have considered here.

\noindent {\bf S2. Astrophysical considerations and implications}

While the detailed parameters of the astrophysical sources at high
redshift are highly uncertain, strong 21-cm absorption is a generic
prediction. A scan through a wide range of currently plausible
astrophysical parameter values\cite{GlobalStudy} (without b-DM
scattering) shows that all models feature an absorption dip during
cosmic dawn\cite{Fur06}, produced (in the direction of increasing
$\nu$) by a fall (i.e., increasing absorption) due to increasing
Lyman-$\alpha$ coupling followed by a rise due to increasing X-ray
heating (or due to reionization in models with late X-ray heating);
the depth of the absorption dip can fall\cite{GlobalStudy} anywhere in
the range $-240$~mK~$<T_{21,{\rm min}} < -25$~mK, and its position in
the range 52~MHz~$< \nu_{\rm min} < 120$~MHz (corresponding to
11~$<z_{\rm min} < 26.5$).

Once b-DM scattering is included, the observed global 21-cm signal is
determined by a complex interplay of this scattering with
astrophysics. For example, in the large allowed region (Fig.~3) of low
$m_X$ and high $\sigma_1$, the initial cooling due to b-DM scattering
can be extremely effective and lead to global 21-cm absorption of
$-1000$~mK or even stronger in the dark ages (though only at very high
redshifts above 100). In these models, the gas is so cold that
Lyman-$\alpha$ coupling is delayed due to low-temperature corrections
(discussed just below), and X-ray heating is also delayed as it
initially must counteract the b-DM cooling. This tends to produce a
relatively deep and wide absorption dip except with particular
astrophysical parameters; this region in parameter space may thus be
disfavored by the data\cite{Judd}, but a clear conclusion requires a
full consideration of the large variety of possible astrophysical
parameters, which we leave for future work. We also note that a very
high $\sigma_1$ would tend to suppress the relative b-DM velocity and
with it the associated fluctuations (discussed in the main text),
though in any case the normal 21-cm fluctuations due to inhomogeneous
galaxy formation would be enhanced in proportion to the (unexpectedly
large) absolute value of the mean global signal.

The observed global 21-cm signal\cite{Judd} implies the first
detection of some of the earliest stars. The location of the peak
absorption at $z \sim 17$ is not especially surprising, but it
significantly narrows down astrophysical parameters that were
previously almost unconstrained. In general, the maximal absorption
corresponds to the late stages of Lyman-$\alpha$ coupling and the
early stages of X-ray heating. The observed timing of these
astrophysical cosmic milestones is easily within the expected range of
astrophysical parameters\cite{GlobalStudy}. Interestingly, the implied
early heating is consistent with early limits from 21-cm observations
of both the global and fluctuation signals\cite{EDGEShi,SARAS,PAPER},
which disfavor strong absorption at low redshifts (as would be
expected in the case of late heating). Indeed, the detected signal
implies that future 21-cm observations should focus on cosmic dawn,
where the 21-cm signal (in terms of both the global signal and the
power spectrum) is likely much stronger than previously expected, and
not on the later era of cosmic reionization, where the signal strength
is likely to fall within the lower part of the previously-expected
range.

An interesting physical detail is that Lyman-$\alpha$ coupling of the
21-cm line is known to become less effective when the gas temperature
is low\cite{ChSh06,Chen,Hirata,PritF,21cmRev2}. In the standard case,
these low-temperature corrections amount at most to a $20\%$ reduction
in the coupling at any redshift\cite{21cmRev2}; with the lower gas
temperatures encountered in the case of significant b-DM scattering,
the low-temperature corrections can reduce the coupling by an order of
magnitude or more, delaying strong Lyman-$\alpha$ coupling and greatly
changing the global 21-cm signal; indeed, in some possible models the
gas temperature gets so low ($<0.1$~K) that these low-temperature
corrections may need to be re-assessed. Another important point is
that the standard expression for the 21-cm signal (eq.~\ref{e:Tb}) is
a linearization assuming a low 21-cm optical depth (a valid assumption
if there is no b-DM scattering), but we encounter high values and thus
always use the more general expression\cite{Madau}. We also note that
even with very cold gas, direct Lyman-$\alpha$ heating can be
neglected since it is very weak\cite{Chen} compared to X-ray heating.

We note that significant b-DM scattering would have another
interesting astrophysical consequence, as it would effect the
formation of the first stars. The lower gas temperature would reduce
the Jeans mass, and the dissipation of the relative b-DM velocity
would reduce its suppression effects on star formation. Both of these
effects would tend to boost star formation relative to the case of no
b-DM scattering, but the impact may be limited since these various
effects on star formation are built up over time; baryon infall into
forming DM halos begins at recombination, and for most DM parameter
values it takes some time until the b-DM scattering has a significant
effect. We have neglected effects on galaxy formation in this work
since they are dwarfed by the direct effect of excess gas cooling on
the 21-cm signal.

\noindent {\bf S3. The range of DM particle masses and b-DM scattering
  cross-sections that can affect cosmic dawn}

The DM parameters that affect 21-cm cosmology are shown in detail in
Fig.~3, but it is important to understand the physics behind the most
important features. In particular, we can easily understand why there
is an upper limit on the DM particle mass that can significantly cool
the cosmic baryons by considering the maximum possible cooling. As
mentioned in section~S1, the baryons thermally decoupled from the CMB
at $z \sim 150$. In the presence of b-DM scattering, by that time the
DM has acquired a non-zero temperature, but the best-case scenario for
maximum cooling of the baryons is that still $T_{\chi} \ll
T_\mathrm{gas}$ at this time. For cooling to occur, the two fluids
must be significantly coupled after the baryons thermally decouple, so
that the baryons share some of their energy with the baryons. The most
that such coupling can achieve, if it is strong, is a thermal
equilibrium at which both the baryons and the DM come to a common
(time-dependent) temperature $T_{\rm fin}$. Then at a given time, the
relation between the baryon temperature $T_\mathrm{gas}$ in the
absence of b-DM scattering and the lowest possible temperature $T_{\rm
  fin}$ with scattering is given by conservation of energy (per unit
volume) as \begin{equation} T_{\rm fin} = T_\mathrm{gas}
  \frac{n_b}{n_b+n_{\chi}} =
  \frac{T_\mathrm{gas}}{1+(\rho_{\chi}/\rho_b)(\mu_b/m_{\chi})} \sim
  \frac{T_\mathrm{gas}}{1+(6~{\rm GeV})/m_{\chi}}\ , \end{equation}
where $n_b$ and $n_{\chi}$ are the number densities of baryons and DM,
respectively, $\rho_b$ and $\rho_{\chi}$ the corresponding (mean)
densities, $\mu_b$ is the mean baryonic mass, and $m_{\chi}$ the mass
of a DM particle. Note that we have neglected here the effect of the
initial b-DM relative velocity, but the associated kinetic energy
would only produce more heating. As an example, in order to reach
$T_{21} = -315$~mK at $z = 17$, the simple estimate in eq.~3 yields a
maximum possible $m_{\chi}$ of 12~GeV. In reality, the cooling never
reaches the best-case scenario assumed in this simple estimate, and we
find an actual maximum mass of 3.5~GeV (see Fig~3). If the observed
peak absorption is determined to be even stronger then this limit
would improve.

There is no lower limit on the DM particle mass that can affect the
21-cm signal, since the cooling rate becomes independent of $m_{\chi}$
when $m_{\chi} \ll \mu_b$. In that limit, the energy lost by a baryon
per collision (at a given b-DM relative velocity) is $\propto
m_{\chi}$, while the scattering rate is $\propto n_{\chi} \sigma_1$,
so the total cooling rate is $\propto \rho_{\chi} \sigma_1$, where
$\rho_{\chi}$ is the (known) mean density of DM. Thus, a significant
interaction requires a minimum $\sigma_1$ that is independent of
$m_{\chi}$ when $m_{\chi} \ll \mu_b$ (Fig.~3).

The dependence of the effectiveness of the baryonic cooling on the
b-DM scattering cross-section is non-trivial. A higher cross-section
means that more of the thermal energy of the baryons can be
transferred to the DM, but on the other hand it also implies that the
DM warms up more early on, before thermal decoupling of the gas from
the CMB, which reduces the ability of the DM to later cool the gas.
There is even a region (e.g., $\sigma_1 = 2 \times 10^{-18}$~cm$^2$
and $m_{\chi} = 100$~GeV; see Fig.~3) where the b-DM interaction
causes a small net baryonic {\it heating}\/ since its main effect is
to transfer kinetic energy from the b-DM relative velocity to the
random gas motions.

\noindent {\bf S4. Comparing cosmic dawn to other limits on b-DM
  interactions}

The comparison between cosmic dawn as a dark matter detector and
constraints from direct detection, accelerators, and various
astrophysical phenomena, is model dependent. We assume here the
$\sigma(v) \propto v^{-4}$ model, in which case the parameter spaces
overlap and other searches may be able to detect or rule out a DM
particle that is relevant for 21-cm observations of cosmic
dawn. However, we note that a more complex interaction, e.g., based on
a bound state or resonance that is significant only at low velocities,
could invalidate any such comparison and make cosmic dawn a unique
probe. Additional model dependence enters in some of the comparisons
which involve also assumptions about DM annihilation or the spin
dependence of the b-DM scattering.

Limits on the $v^{-4}$ model have been previously
derived\cite{McDermott}, with the strongest limits based on CMB
observations (plus a slight improvement from including clustering
based on Lyman-$\alpha$ forest data)\cite{Marc}. The 95\% confidence
limit equivalent to $\sigma_1 < 2 \times 10^{-19} (m_{\chi}/{\rm
  GeV})$~cm$^2$ was, however, derived only for $m_{\chi} \gg
m_H$. This calculation must be re-done for lower-mass $m_{\chi}$,
along with a proper inclusion of the spatial variation of the b-DM
relative velocity, which would introduce a CMB pattern that may be
partially degenerate with the standard one. We can approximately
estimate the correction for lower DM mass; in the limit of strong
coupling (so that $T_{\chi} = T_\mathrm{gas}$), including the
contribution (neglected in the above limit) of the DM to the relative
thermal velocity, and assuming that the limit $\propto v^{-4}$, gives
a modified limit of $\sigma_1 < 2 \times 10^{-19} (m_{\chi}/{\rm GeV})
[1+(\mu_b/m_{\chi})]^2$~cm$^2$. Note, though, that if the coupling is
not strong, then $T_{\chi} < T_\mathrm{gas}$ and the correction factor
is smaller. We conclude that CMB limits may complement the 21-cm
signal by giving significant upper limits on $\sigma_1$, but these
limits must be carefully re-calculated. There is also a limit on b-DM
scattering from spectral distortions of the CMB\cite{spectral}, but
those distortions occur at rather high redshifts (and thus high
velocities), so have not been considered for a cross-section that
peaks at low velocities such as the $\sigma(v) \propto v^{-4}$ model.

A speculative possibility to be checked is that between
matter-radiation equality and recombination, a period when
fluctuations normally grow in the DM but not the baryons, the b-DM
coupling might suppress the growth of fluctuations in the DM, and
perhaps this effect can be consistent with CMB observations yet still
significant for the power spectrum at low redshift.

It is interesting to consider the specific DM millicharge model, which
as mentioned above naturally yields a $v^{-4}$ cross-section. In this
model, after recombination the cross-section effectively drops in
proportion to the residual proton fraction. Thus, the effect of DM
scattering on baryon cooling is suppressed relative to the effect on
the CMB by a factor of $\sim 5000$. In this case, the CMB limit (with
the above estimated correction) would rule out a significant effect on
the 21-cm signal at cosmic dawn unless $m_{\chi} < 60$~MeV.

A different limit on b-DM interactions comes from experiments that
attempt to directly detect DM scattering with target nuclei in the
lab. Assuming typical Milky Way halo speeds of order the rotation
velocity of $\sim 200$~km/s, the minimum $\sigma_1 > 10^{-21}$~cm$^2$
required for a cosmic 21-cm effect (Fig.~3) translates (in a $v^{-4}$
model) to $\sigma(200\ {\rm km/s}) > 6 \times 10^{-31}$~cm$^2$. This
falls right in the range of cross-sections that are hard to probe with
underground detection experiments, since at such cross-sections the DM
particles are expected to lose most of their energy in the Earth's
crust before reaching the detector\cite{Farrar}. The CRESST
underground experiment is most relevant to the 21-cm parameter region,
though it constrains only relatively high masses; at $m_\chi = 1 -
5$~GeV, it rules out (assuming spin-independent interactions
throughout this discussion) values of $\sigma(200\ {\rm km/s})$
between $10^{-37}$~cm$^2$ and $2-3 \times 10^{-31}$~cm$^2$ (the upper
limit varies somewhat with $m_\chi$), though the limit might be
stronger when re-calculated for a $v^{-4}$ model (since the particles
slow down as they scatter within the Earth). Experiments above the
Earth's surface thus have an advantage. A 1987 balloon experiment
rules out $m_\chi > 2-3$~GeV (the precise limit depends on
uncertainties in the DM halo velocity
distribution)\cite{Balloon,Farrar}. The rocket-based X-ray quantum
calorimetry (XQC) experiment excludes\cite{XQC} $\sigma(200\ {\rm
  km/s}) > 1 \times 10^{-29}$~cm$^2$ for $m_{\chi} > 0.5$~GeV. The
limits from all such experiments on lower DM particle masses are quite
weak (although this could change with new
techniques\cite{Budnik,DM2017}). A much stronger limit comes from the
flip-side of the Earth interactions mentioned above. The scatter of
the DM particles within the Earth would heat it too
strongly\cite{HeatOld,Heat} unless $\sigma(200\ {\rm km/s})<
10^{-32}\,({\rm GeV}/m_\chi)$~cm$^2$, valid for $m_\chi$ of a few GeV
and below; this implies that a significant 21-cm effect at cosmic dawn
requires $m_\chi < 15$~MeV. However, the Earth-heating constraint
relies on some assumptions regarding DM annihilation. Also, this and
all the above direct detection limits on $\sigma$ are weaker by 3--4
orders of magnitude for spin-dependent interactions\cite{Heat,Farrar},
while the cosmic scattering with hydrogen would remain just as strong
for such an interaction given hydrogen's nuclear spin of $I=1/2$ (note
that $^4$He would not contribute in that case).

In high-energy particle accelerators, since the collision energies are
typically much greater than a GeV, we assume that the relevant
cross-section is $\sigma_c$. Accelerators may probe some of the
parameter space that is relevant for 21-cm cosmology, but comparing
the limits depends strongly on the precise interaction type and nature
of the DM particle\cite{DM2017,SHiP}. The proposed SHiP experiment at
CERN has been motivated by the many possible physical mechanisms for
producing very weakly interacting DM particles in the MeV-GeV mass
range\cite{SHiP}; cosmology may now provide additional impetus.

Astrophysical constraints on b-DM interactions are generally weaker
than those we have considered\cite{HeatOld}. The most significant
limit, from cosmic rays, is often quoted as $\sigma(200\ {\rm km/s}) <
8\times 10^{-27} (m_\chi/{\rm GeV})$~cm$^2$, but this is only valid
for large $m_\chi$ (greater than $\sim 100$~GeV), and the limits on DM
particle masses in our range of interest are far weaker\cite{Cyburt}.

We again emphasize that we have assumed throughout this section a
$v^{-4}$ dependence of the b-DM scattering cross-section, but any
modification of this would have a major effect on the above
comparisons of the cosmological signal with the various limits on b-DM
interactions (as would other extensions of the parameter space such as
allowing only a fraction of the DM to scatter with baryons); also,
previous constraints have often been derived for a
velocity-independent cross-section and must be carefully re-assessed
for the case of a strong velocity dependence. We also note that the
constraints on the DM particle from it having the required relic
cosmic density are model-dependent, as they depend on the annihilation
cross-section (for thermal production) or the detailed production
mechanism (for non-thermal production). As a simple example, if we
assume thermal production, an annihilation cross-section as expected
for the weak interaction, and a similar b-DM cross-section at
relativistic velocities (so that $\sigma_c \sim 10^{-36}$~cm$^{-2}$),
then the 21-cm signal suggests something closer to a $v^{-3}$
dependence rather than $v^{-4}$.

Finally we note that after this {\it Letter}\/ was submitted, limits
on b-DM scattering were derived\cite{ML2017} from low redshift
observations of the temperature of the intergalactic medium, based on
the Lyman-$\alpha$ forest at $z \sim 5$. These limits are effectively
at relative velocities of $\sim 10$~km/s, which is similar to the CMB
limits discussed above. The derived upper limit of $\sigma_c < 3
\times 10^{-38}$~cm$^2$ for a $v^{-4}$ model (for $m_\chi << 1$~GeV)
is about five orders of magnitude above the minimum cross-section that
can be probed at cosmic dawn, and is stronger than the above CMB limit
only for $m_\chi$ below $\sim 1$~MeV (although, as we noted, the CMB
limit must be carefully revised). Moreover, the validity of this
precise low-redshift limit is unclear since it depends on the history
of photoheating of the intergalactic medium; the latter depends on the
spatial and temporal distribution of the spectrum of ionizing sources,
the distribution of Lyman-limit absorbers, and the after-affects of
inhomogeneous reionization, all of which are incompletely known. In
the low-redshift probe, astrophysical heating is partly degenerate
with b-DM interactions, and there is no unambiguous sign of DM similar
to the excess absorption signal during cosmic dawn.

\end{document}